\documentclass[twocolumn]{aastex61}
\usepackage{amsmath}

\shorttitle{
Tidally induced bars from galactic flybys}
\shortauthors{E. L. {\L}okas}

\begin{document}

\title{Formation of tidally induced bars in galactic flybys:\\ prograde versus retrograde encounters}

\author{Ewa L. {\L}okas}
\affil{Nicolaus Copernicus Astronomical Center, Polish Academy of Sciences,
Bartycka 18, 00-716 Warsaw, Poland}

\begin{abstract}
Bars in disky galaxies can be formed by interactions with other systems, including
those of comparable mass. It has long been established that the effect of such interactions on galaxy morphology
depends strongly on the orbital configuration, in particular the orientation of the intrinsic spin of the
galactic disk with respect to its orbital angular momentum. Prograde encounters modify the morphology
strongly, including the formation of tidally induced bars, while retrograde flybys should
have little effect on morphology. Recent works on the subject reached conflicting conclusions, one using the
impulse approximation and claiming no dependence on this angle in the properties of tidal bars. To resolve the
controversy, we performed self-consistent $N$-body simulations of hyperbolic encounters between two identical Milky
Way-like galaxies assuming different velocities and impact parameters, with one of the galaxies on a prograde and the
other on a retrograde orbit. The galaxies were initially composed of an exponential stellar disk and an NFW dark halo,
and were stable against bar formation in isolation for 3 Gyr. We find that strong tidally
induced bars form only in galaxies on prograde orbits. For smaller impact parameters and lower relative velocities
the bars are stronger and have lower pattern speeds. Stronger bars undergo extended periods of buckling instability
that thicken their vertical structure. The encounters also lead to the formation of two-armed spirals with
strength inversely proportional to the strength of the bars. We conclude that proper modeling of prograde and
retrograde encounters cannot rely on the simplest impulse approximation.
\end{abstract}

\keywords{
galaxies: clusters: general --- galaxies: evolution --- galaxies: fundamental parameters --- galaxies: interactions
--- galaxies: kinematics and dynamics --- galaxies: structure }

\section{Introduction}

A significant fraction of late-type galaxies in the Universe possess bars \citep[see e.g.][and references
therein]{Buta2015}. These prominent morphological features can originate from instabilities in
self-gravitating axisymmetric disks \citep[for a review see][]{Athanassoula2013} and their evolution depends on the
initial structural parameters of the disk and the dark matter halo of the galaxy. Another formation channel of
the bars involves interactions of a galaxy with perturbers of different masses \citep{Noguchi1996, Miwa1998}. These may
include tidal effects of a neighboring bigger structure, as in the case of a satellite galaxy orbiting a bigger host
\citep{Lokas2014, Lokas2015, Gajda2017} or a normal-size galaxy orbiting a cluster \citep{Mastropietro2005, Lokas2016}.
Bars can also be induced by satellites infalling into a bigger galaxy \citep{Mayer2004}. A special case among these are
interactions between objects of similar mass, recently considered by \citet{Lang2014} and \citet{Martinez2017}.

The key parameter controlling the outcome of the interaction, in addition to the strength of the tidal force, seems to
be the angle between the spin of the galaxy disk and its orbital angular momentum with respect to the perturber. Ever
since the seminal work of \citet{Toomre1972} we know that the prograde or direct encounters (when the two angular
momenta are aligned) have much more dramatic effect on the galactic structure than retrograde ones (when the angular
momenta point in opposite directions). In particular, \citet{Toomre1972} demonstrated using a restricted three-body
approach that prograde encounters lead to the formation of long and narrow tidal arms while in the retrograde cases the
disks remain rather unaffected. More recent work based on the improved impulse approximation \citep{DOnghia2010}
confirmed that this is due to a broad resonance between the angular velocities of stars in the galaxy and their angular
velocities on the orbit around the perturber. In agreement with these theoretical predictions, \citet{Lokas2015} and
\citet{Lokas2016} found using self-consistent $N$-body simulations that tidally induced bars form in galaxies orbiting
a bigger host if they are on prograde orbits, but not if they are on retrograde ones.

Although we expect that the general effect should not depend on the relative masses of the galaxy and the perturber,
the picture has been made less clear in recent studies of equal-mass encounters between galaxies. While \citet{Lang2014}
found that the tidally induced bars form only in prograde encounters and not the retrograde ones, \citet{Martinez2017}
claimed the results to be independent of the inclination: their bars formed in retrograde encounters turned out to be
similar in strength to those forming in direct ones.
However, \citet{Martinez2017} used the impulse approximation instead of fully self-consistent simulations that
must have affected their results.
Although they attempted to test the robustness of the impulse approximation by full simulations, they
did so for cases that are not representative of the whole study and therefore cannot be considered as convincing.

To resolve the controversy, in this paper we revisit the problem of the formation of tidally induced bars in equal-mass
flybys. For this purpose we use self-consistent $N$-body simulations of encounters between two identical galaxies
similar to the Milky Way. Each galaxy is initially in equilibrium and its structural parameters are such that it is
stable against bar formation in isolation for a few Gyr. The galaxies are placed on hyperbolic orbits so
that one has an exactly prograde and one exactly retrograde orientation while both act as each other's perturber.
In section~2 we provide the details of the simulations. Section~3 describes the formation and properties of the bars
occurring in the prograde encounters in contrast to very weak distortions present in the retrograde cases. In
section~4 we look in more detail into the origin of the difference between the prograde and retrograde interactions and
the conclusions follow in section~5.

\section{The simulations}

We used a self-consistent $N$-body realization of a Milky
Way-like galaxy close to equilibrium constructed using procedures described in \citet{Widrow2005}
and \citet{Widrow2008}. The galaxy had two components: an NFW \citep{Navarro1997} dark matter halo and an exponential
stellar disk, but no classical bulge. Each of the two components was made of $10^6$ particles. The structural parameters
were similar to the model MWb of \citet{Widrow2005}: the dark matter halo had a virial mass $M_{\rm H} = 7.7 \times
10^{11}$ M$_{\odot}$ and concentration $c=27$. The disk had a mass $M_{\rm D} = 3.4 \times 10^{10}$ M$_{\odot}$, the
scale-length $R_{\rm D} = 2.82$ kpc and thickness $z_{\rm D} = 0.44$ kpc.

The model of the galaxy is thus the same as
the one used by \citet{Lokas2016} and \citet{Semczuk2017} to study the formation of tidally induced bars and spiral
arms in galaxies orbiting a Virgo-like cluster. The minimum value of the Toomre parameter of this realization was
$Q=2.1$ and the model was demonstrated to be stable against bar formation for 3 Gyr \citep{Lokas2016, Semczuk2017}.
Note that although we could have made the galaxy more stable by increasing the Toomre parameter even
more, we wanted to keep the model as similar as possible to the well-studied case of the Milky Way. In addition, the
bar that starts to form in this galaxy after 3 Gyr of evolution is very weak and needs another 3 Gyr to fully develop.

\begin{figure}
\gridline{\fig{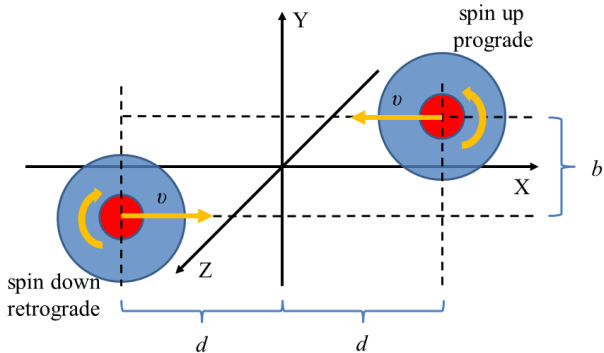}{0.45\textwidth}{}}
\vspace{-0.03\textwidth}
\caption{Schematic view of the initial configuration of the flybys. The two disky Milky Way-like galaxies (with red
circles depicting disks and blue ones dark matter halos)
are placed at the distance $d$ from the $Y$ axis and $b/2$ from the $X$ axis of the coordinate system. They are
assigned velocities $v$ with opposite directions along the $X$ axis. The spin of the galaxy disk on the right (left) is
up (down), so it is prograde (retrograde) with respect to the orbital angular momentum. The values of parameters
$d$, $b$ and $v$ are listed in columns 2-4 of Table~\ref{configuration}.}
\label{configuration}
\end{figure}

Our flyby configurations are such that in each simulation one galaxy with above parameters is on a prograde orbit,
while the other identical galaxy has a retrograde orientation. In this way, the galaxies act as each other's
perturber and we can study the effect of prograde and retrograde encounters using the same simulation. The
configuration used for all simulations is shown in Figure~\ref{configuration}. The collision takes place in the
$XY$ plane of the simulation box. The galaxy on the right has a disk coplanar with the orbit with the spin up and is
moving initially to the left with velocity $v$, aligned with the $X$ axis. It is placed at a distance $d$ from the $Y$
axis and $b/2$ from the $X$ axis (so that $b$ is the impact parameter). The second galaxy, on the left, has its disk
also coplanar with the orbit but with the spin oriented down and is initially moving with the same velocity $v$, but in
the opposite direction, to the right. Its initial position is exactly symmetric with respect to the first galaxy. Note
that the velocity values are given with respect to the coordinate system of the simulation box, so the relative
velocity of the two galaxies is $2v$.

We considered four different combinations of the encounter parameters. In the first simulation, which we
refer to as S1, we used $d=500$ kpc, $b=50$ kpc and $v=500$ km s$^{-1}$. This choice of velocities was made in order to
mimic the lower end of orbital velocities characteristic of galaxy clusters, since this is where we expect such
flybys to take place. As shown in the middle panel of fig. 2 in \citet{Lokas2016}, the velocities of galaxies on
typical orbits in a medium-size cluster vary between 300 and 1900 km s$^{-1}$. These parameters lead to the closest
passage occurring around 1 Gyr after the start of the simulation with the pericenter approximately equal to the impact
parameter $b$. Note also, that $b=50$ kpc corresponds to about twice the extent of the galactic disk, so the disks
barely touch during the closest approach.

\begin{table}
\begin{center}
\caption{Configuration details of the simulations}
\begin{tabular}{lccccl}
\hline
\hline
Simulation  & $d$   &  $b$  &  $v$ &  $ S $ & Line color\\
            & [kpc] & [kpc] &[km s$^{-1}$]&  &\\
\hline
\ \ \ \ S0 &  $-$  & $-$   & $-$  &  $-$    & \ \ black   \\

\ \ \ \ S1 &  500  & 50   & 500   &  0.02  & \ \ blue    \\

\ \ \ \ S2 &  500  & 25   & 500   &  0.07  & \ \ cyan    \\

\ \ \ \ S3 &  350  & 25   & 350   &  0.15  & \ \ green   \\

\ \ \ \ S4 &  250  & 25   & 250   &  0.26  & \ \ red     \\
\hline
\label{tabconfiguration}
\end{tabular}
\end{center}
\end{table}

To increase the strength of the interaction, in simulation S2 we decreased the impact parameter to $b=25$ kpc. In this
case, since the velocities remain large, the trajectories are still not strongly affected before the encounter and
the pericenter of the flyby is again close to $b$. Then, in simulations S3 and S4 we keep the smaller impact
parameter of $b=25$ kpc and additionally decrease the velocity down to $v=350$ km s$^{-1}$ and $v=250$ km s$^{-1}$,
respectively, so that they are now more typical of galaxy groups than clusters. In order to have the encounter
happening again around 1 Gyr after the start of the simulation we also decrease the initial distance between the
galaxies by the same factor, to have $d=350$ kpc and $d=250$ kpc. In these cases, the trajectories are affected before
the encounter and the actual pericenters are 21 and 17 kpc, respectively, significantly smaller than $b$. For reference,
we also include in our comparisons the simulation of the evolution of the same galaxy in isolation and refer to it as
S0.

The parameters of the encounters are listed in columns 2-4 of Table~\ref{tabconfiguration}. In the fifth
column we list the corresponding values of the dimensionless tidal strength parameter $S$ of \citet{Elmegreen1991}. We
note that \citet{Elmegreen1991} estimated the threshold for permanent bar formation in their simulations to be $S>0.04$
so we can expect bars to form in our encounters S2-S4. The last column of the Table lists the line colors with which
the results for the different simulations will be shown throughout the paper.

The evolution of the systems was followed for 3 Gyr with the $N$-body code GADGET-2 \citep{Springel2001, Springel2005}
saving outputs every 0.05 Gyr. The adopted softening scales were $\epsilon_{\rm D} = 0.1$ kpc and $\epsilon_{\rm H} =
0.7$ kpc for the disk and halo of the galaxies, respectively.

\begin{figure}
\gridline{\fig{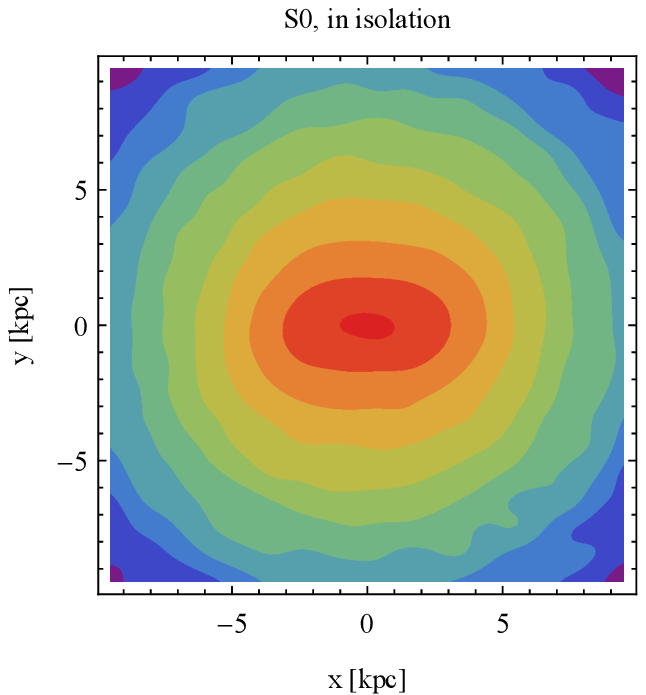}{0.22\textwidth}{}
	  \hspace{0.032\textwidth}
	  \fig{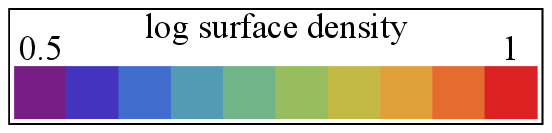}{0.17\textwidth}{}
	  \hspace{0.005\textwidth}
	  }
\vspace{-0.05\textwidth}
\gridline{\fig{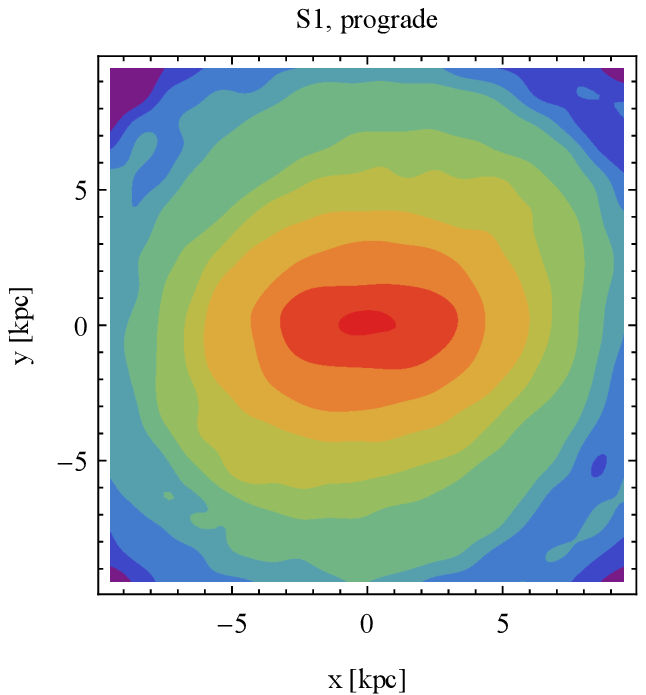}{0.22\textwidth}{}
          \fig{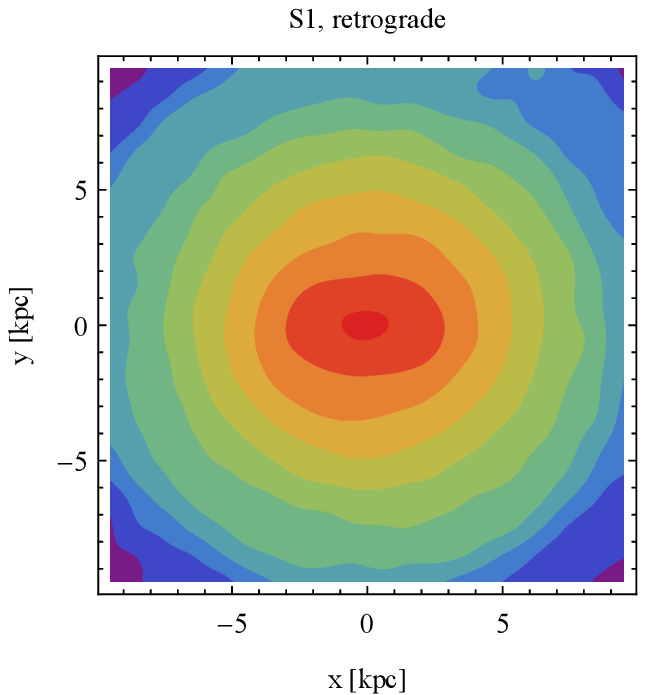}{0.22\textwidth}{}
	  }
\vspace{-0.05\textwidth}
\gridline{\fig{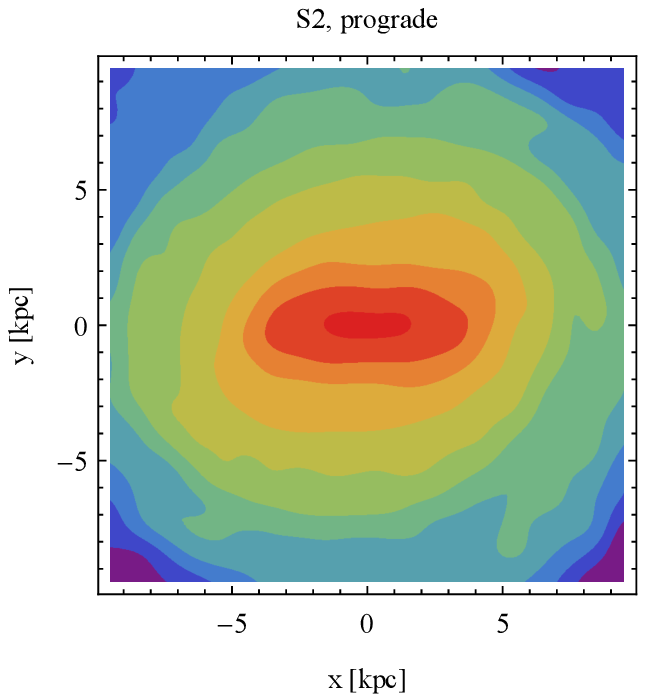}{0.22\textwidth}{}
          \fig{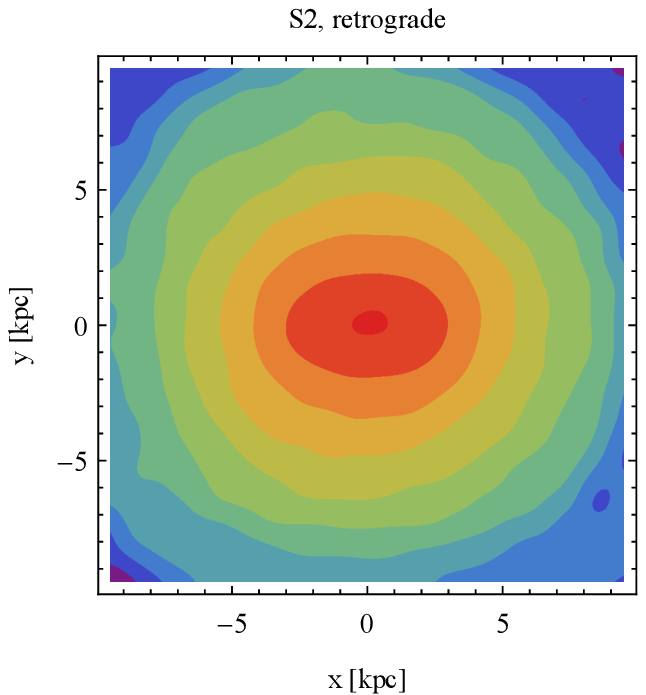}{0.22\textwidth}{}
          }
\vspace{-0.05\textwidth}
\gridline{\fig{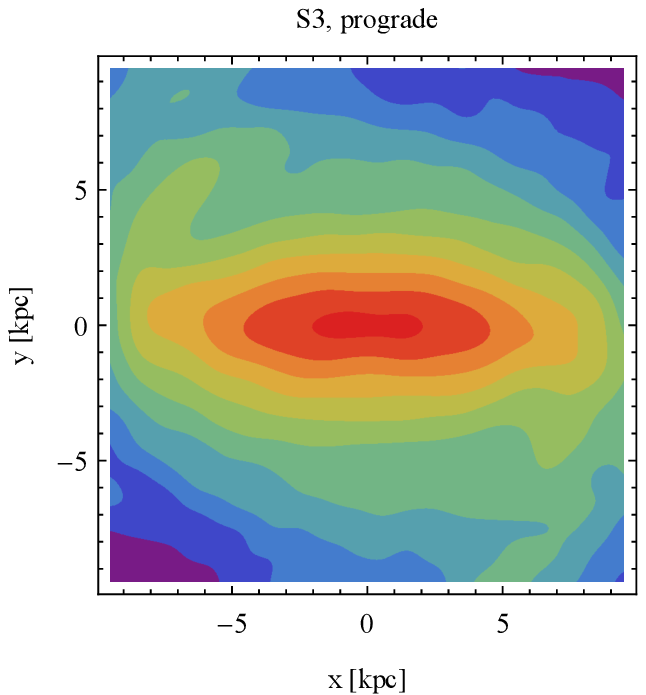}{0.22\textwidth}{}
          \fig{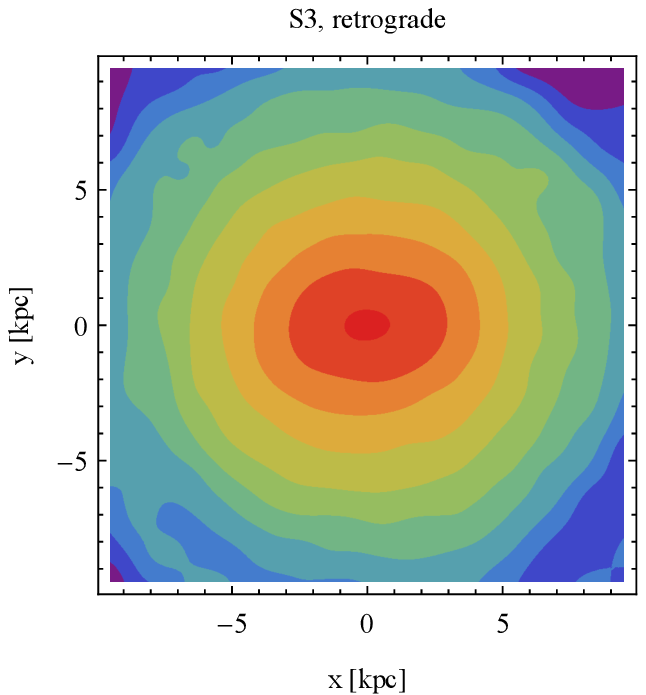}{0.22\textwidth}{}
          }
\vspace{-0.05\textwidth}
\gridline{\fig{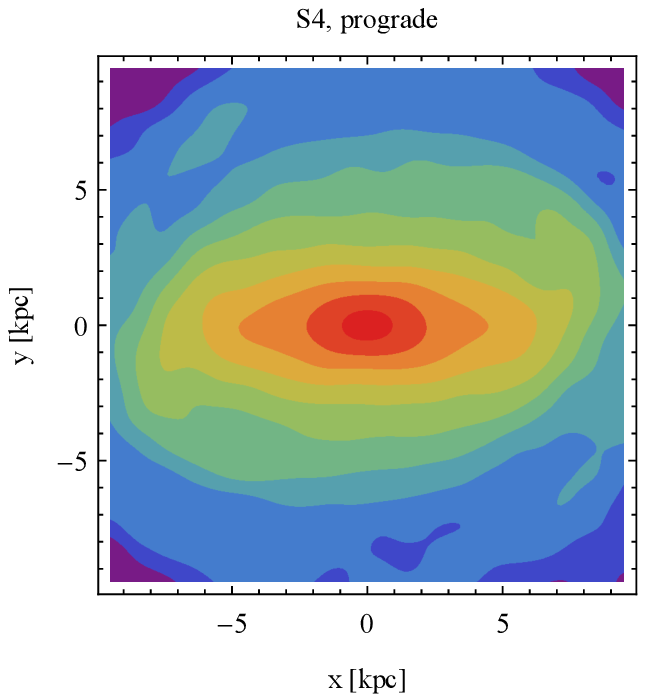}{0.22\textwidth}{}
          \fig{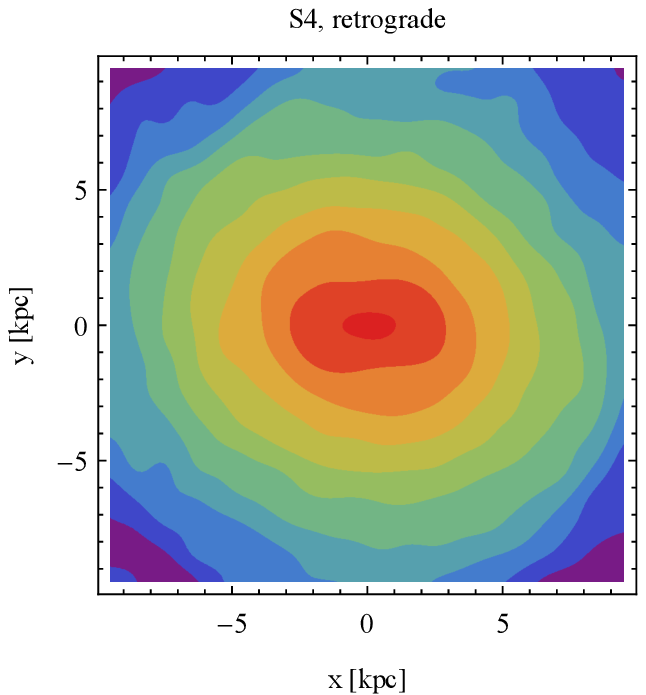}{0.22\textwidth}{}
          }
\vspace{-0.04\textwidth}
\caption{
Surface density distributions of the stars in the galaxies at the end of the evolution in the face-on view.
}
\label{surdenrot}
\end{figure}

\begin{figure*}
\gridline{
	\fig{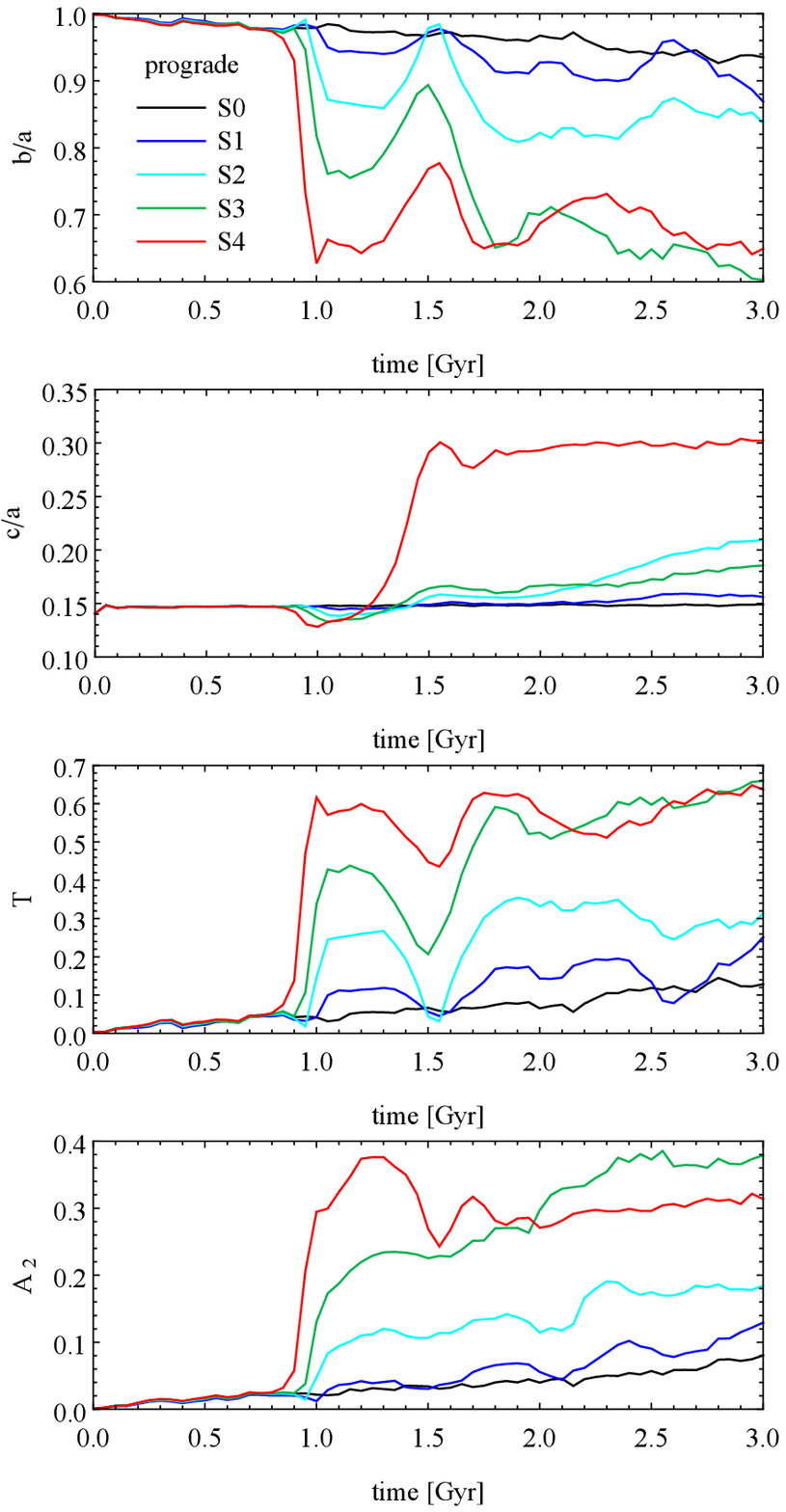}{0.46\textwidth}{}
	\fig{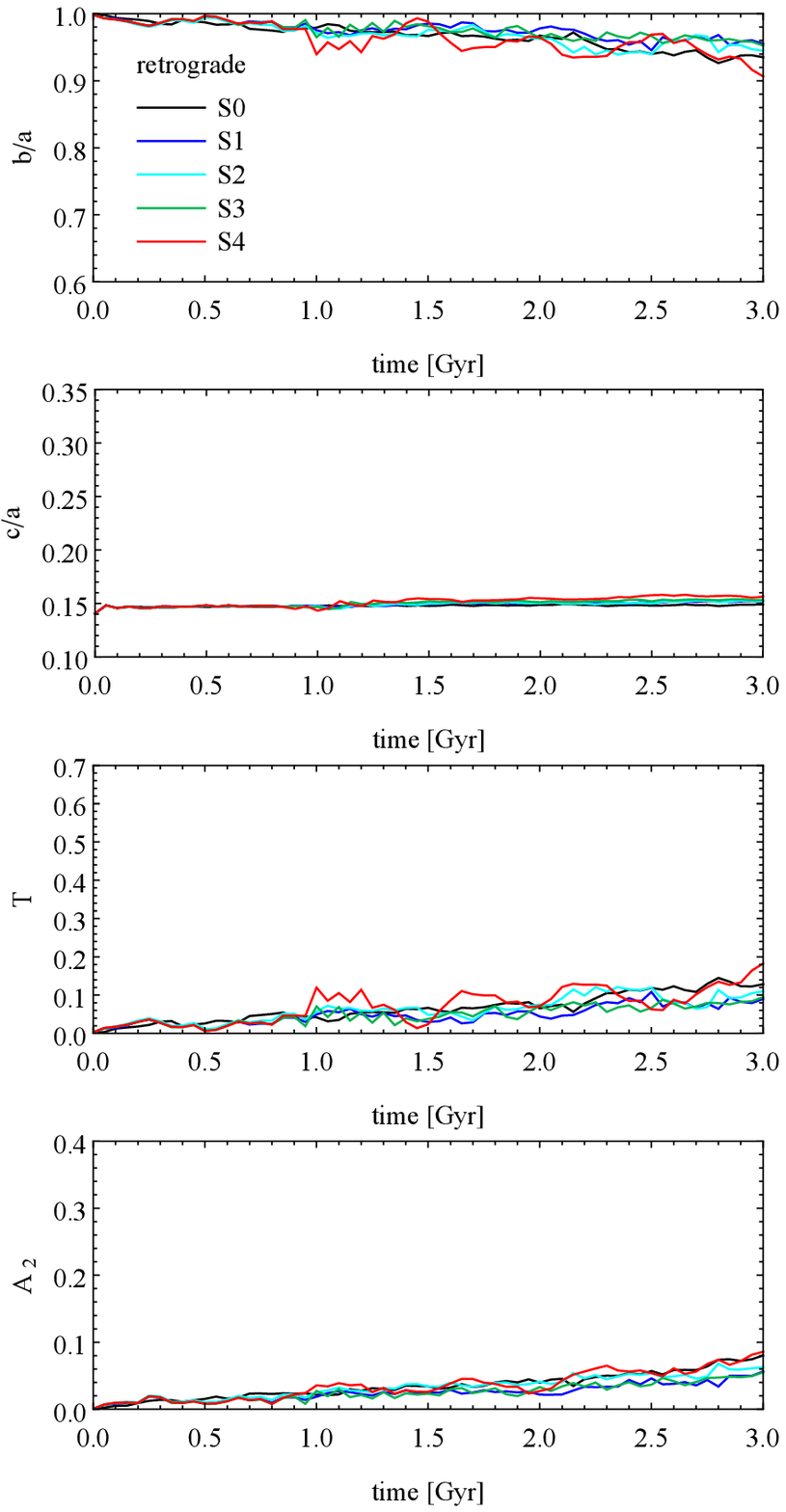}{0.46\textwidth}{}
	}
\vspace{-0.03\textwidth}
\caption{The evolution of different measures of the shape of the stellar component in time in simulations S0-S4
for galaxies on prograde (left column) and retrograde (right column) orbits. Simulations
with different initial configurations are denoted with different colors, as indicated by the legend in the upper panels.
The panels from top to bottom show respectively the evolution of the axis ratios $b/a$ and $c/a$, the triaxiality
parameter $T$ and the bar mode $A_2$. The measurements were performed for stars within 7 kpc from the center of the
galaxy.}
\label{shape}
\end{figure*}

\begin{figure*}
\gridline{
          \fig{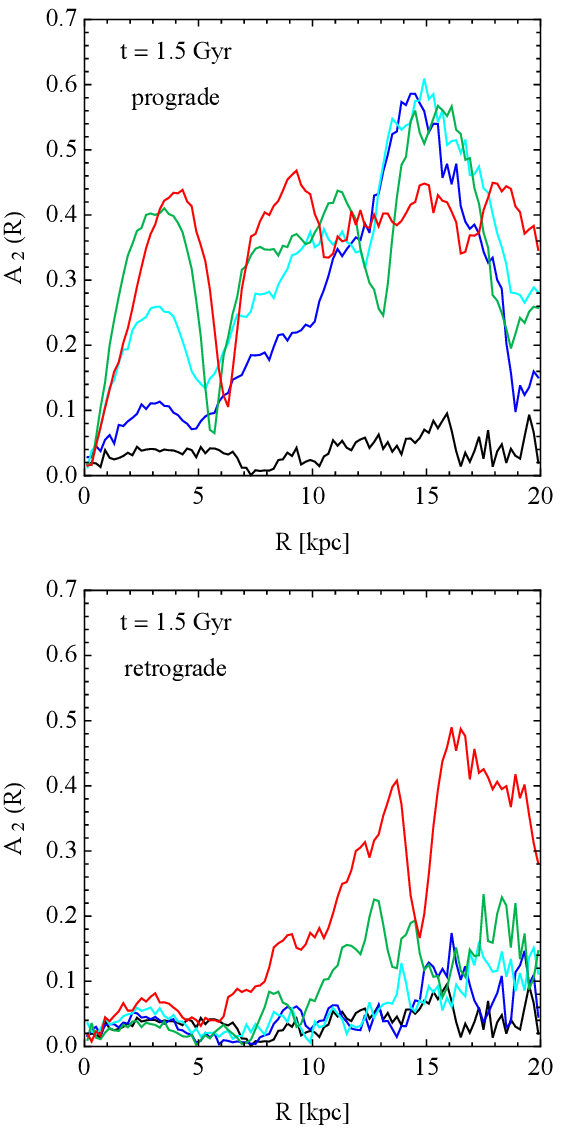}{0.31\textwidth}{}
          \fig{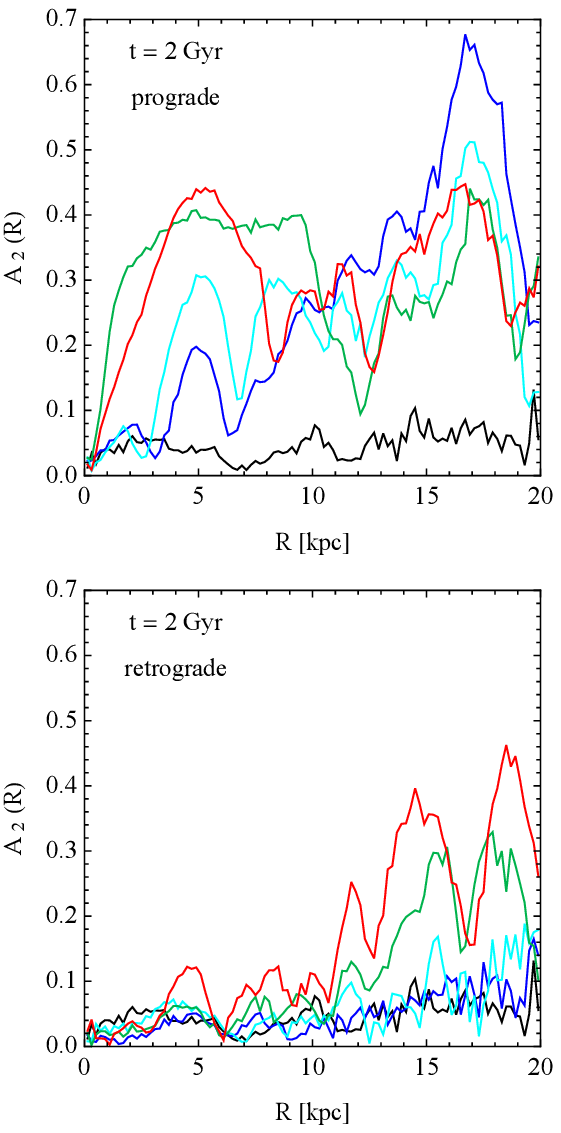}{0.31\textwidth}{}
          \fig{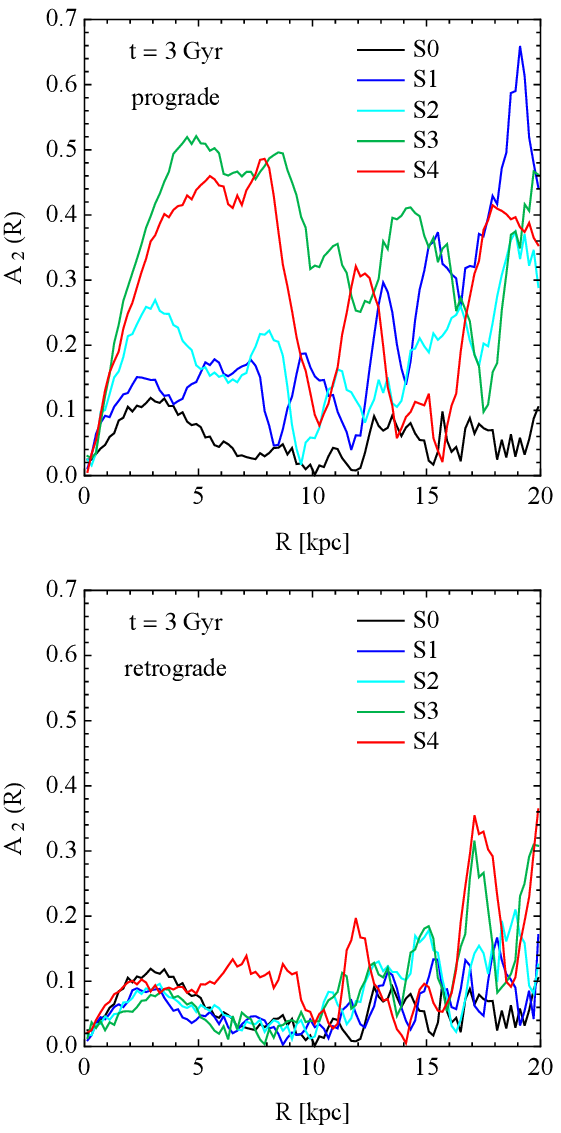}{0.31\textwidth}{}
	  }
\vspace{-0.03\textwidth}
\caption{The profiles of the bar mode $A_2(R)$ for galaxies on prograde (upper panels) and retrograde (lower panels)
orbits at different times $t=1.5$, 2 and 3 Gyr from the start of the simulation (from left to right). Simulations
with different initial configurations are denoted with different colors, as indicated by the legend in the right
panels.}
\label{a2profiles}
\end{figure*}

\begin{figure}
\gridline{\fig{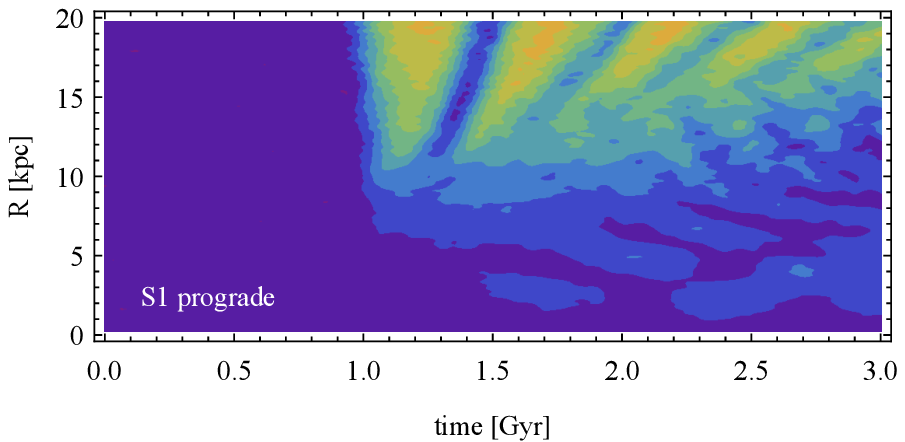}{0.46\textwidth}{}}
\vspace{-0.05\textwidth}
\gridline{\fig{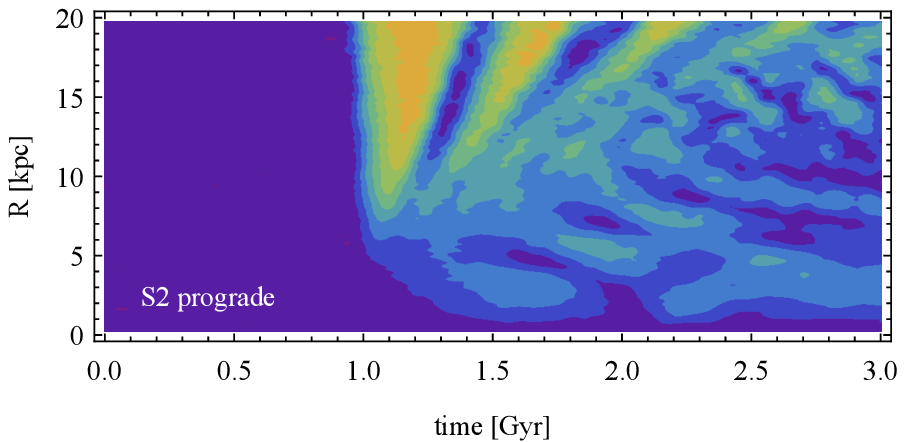}{0.46\textwidth}{}}
\vspace{-0.05\textwidth}
\gridline{\fig{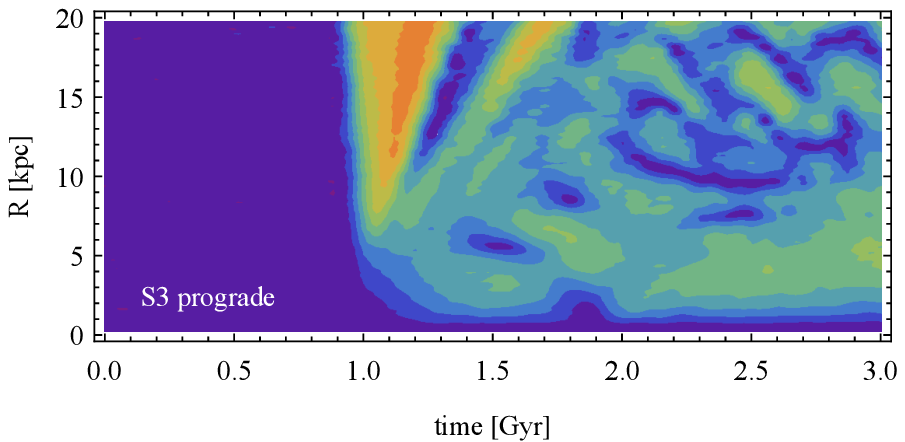}{0.46\textwidth}{}}
\vspace{-0.05\textwidth}
\gridline{\fig{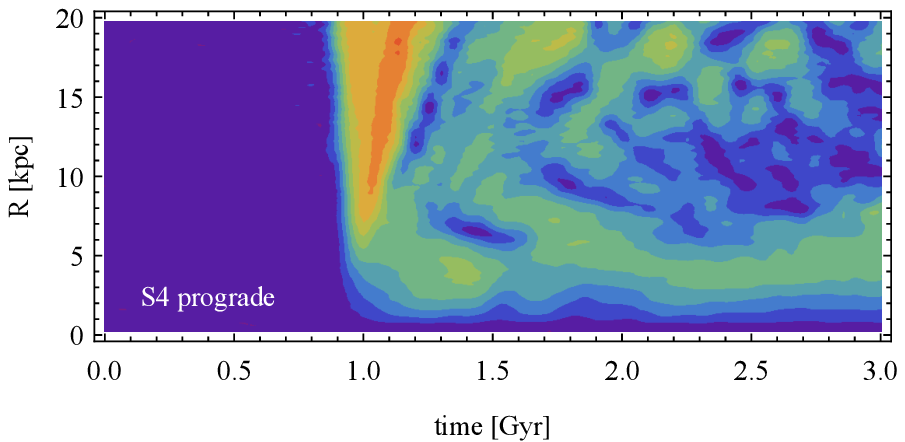}{0.46\textwidth}{}}
\vspace{-0.03\textwidth}
\gridline{\hspace{0.04\textwidth}
	  \fig{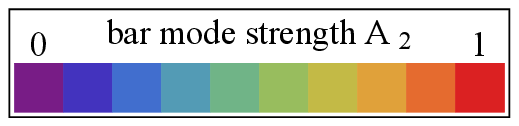}{0.2\textwidth}{}}
\vspace{-0.03\textwidth}
\caption{The evolution of the profile of the bar mode $A_2(R)$ in time for galaxies on prograde orbits in simulations
S1-S4 (from the top to the bottom panel).}
\label{a2modestime}
\end{figure}

\section{Formation and evolution of the bars}

We find that tidally induced bars of significant strength form only in galaxies on prograde orbits in our simulated
flybys. Figure~\ref{surdenrot} shows the images of the galaxies after 3 Gyr of evolution in
terms of the projected surface density distribution of the stellar component in the face-on view, i.e. along the
shortest axis. The galaxies were rotated so that the bar is aligned with the $x$ axis and the shortest axis is along
$z$. The upper panel shows the image for the galaxy evolved in isolation (S0). The lower rows show the distributions
for our four simulations of flybys S1-S4 for prograde (left column) and retrograde (right column) cases.
The measurements were normalized to the maximum value for each galaxy which is largest for S4 and equal to
$\Sigma_{\rm max} = 8.6 \times 10^8$ M$_\odot$ kpc$^{-2}$. Contours are equally spaced in $\log \Sigma$ with $\Delta
\log \Sigma = 0.05$.

The galaxy evolving in isolation shows a slight distortion in the form of a very weak bar that is just starting
to form. For the interacting cases, the bar is weakest in simulation S1 with the larger velocity and impact parameter,
but still a little stronger than in the isolated case. The bar has intermediate strength for S2 with a smaller impact
parameter, and is quite strong for S3 and S4 that had smaller both the velocity and the impact parameter. We note that
it is the bar in S3 and not S4 that is strongest and longest, as will be confirmed by the following measurements. This
suggests that there is no monotonic relation between the strength of the tidal force and the strength of the bar
formed, a tendency that was also noticed in the case of tidally induced bars in dwarfs orbiting the Milky Way by
\citet{Gajda2017} and in a recent study of prograde interactions by \citet{Pettitt2018}. The reason for this
behavior in the particular case of flybys will be discussed below.

The evolution of the shape of the stellar component in time is illustrated in Figure~\ref{shape} for prograde
cases in the left column and the retrograde ones in the right. The two top panels of each column show the axis ratios
$b/a$ and $c/a$, where $a$, $b$ and $c$ are the lengths of the longest, intermediate and shortest axis, respectively.
The third panels plot the triaxiality parameter $T = [1-(b/a)^2]/[1-(c/a)^2]$ and the lowest panels the bar mode $A_2$,
i.e. the module of the Fourier $m=2$ mode of the surface distribution of the stars. All measurements were done using
stars within 7 kpc from the center of the galaxy, which roughly corresponds to the maximum length of the bars (see
below). Lines of different colors were used to plot results for simulation S1 (blue), S2 (cyan), S3 (green) and S4
(red). For comparison, with black lines we also add results for the galaxy evolved in isolation, S0. Clearly, for
prograde galaxies up to 2 Gyr there is a systematic growth of the bar strength and prolateness of the stellar
distribution from S0 to S4, that is with the strength of the interaction. At 2 Gyr the growth of the bar in S3
overcomes that in S4 and this bar ends up strongest in the end. The analogous inner properties of galaxies on
retrograde orbits (right column plots) are very similar to the results for the galaxy in isolation, i.e. they show very
little evolution in time.

The strength of the bar is best characterized by the profile of the module of the Fourier $m=2$ mode of the surface
distribution of the stars. We calculated this parameter for each output of the simulations projecting all stars along
the shortest axis of the stellar component in different radial bins up to the radius of 20 kpc. The results are shown
in Figure~\ref{a2profiles} for outputs corresponding to different times $t=1.5$, 2 and 3 Gyr from the beginning of the
simulation (columns, from the left to the right). In each column the upper panel shows the results for the galaxy on
the prograde orbit, while the lower for the galaxy on the retrograde one.

Comparing the upper panels (prograde) with the lower ones (retrograde) we confirm that in the prograde cases bars of
significant strength form, while in the retrograde cases the $A_2(R)$ profiles within the inner 5 kpc are below 0.1 and
remain on the same level as the bar mode values for the model in isolation. In addition, they are similar for all
retrograde flyby parameters. For the prograde cases the bar mode values vary strongly depending on the force of the
tidal interaction, confirming the impression from the surface density distributions and global shape measurements
discussed above. While for the strongest interactions in S3 and S4 the bar mode is rather high, for the weaker one in
S2 it is significantly lower and for S1 only a little higher at the end than the one for the evolution in isolation.

A more complete view of the bar mode evolution in the prograde cases, taking into account both its radial and time
dependence, can be seen in the maps of Figure~\ref{a2modestime}. In all simulations, the stellar component becomes
strongly elongated right after the time of the closest approach to the second galaxy, around 1 Gyr from the start of
the simulation, as indicated by the high $A_2$ values in the outer parts of the galaxy. This elongation is strongest
for S4 since the encounter is closest in this case and lasts for a longest time because of the lowest relative velocity
of the galaxies. Some time after the encounter, all galaxies show an increase of $A_2$ also in the inner part, at radii
$0 < R < 5$ kpc, signifying the formation of the bar. This again occurs faster and reaches higher $A_2$ values for the
strongest interaction in S4. Only after 2 Gyr of evolution the bar in S3 starts to grow faster and is
strongest and longest in the end. An analogous map for the simulation S0, of the galaxy in isolation, can be found in
the lowest panel of fig. 8 in \citet{Lokas2016}. Let us recall that in this case the bar starts to form no earlier than
3 Gyr, i.e. it crosses the threshold of $A_2(R)>0.1$ for the first time at $t=2.8$ Gyr.

Interestingly, around 2 Gyr from the start of the simulations, the weaker bars formed in (prograde) simulations S1 and
S2 seem to dissolve. Indeed, comparing the upper left and upper middle panel of Figure~\ref{a2profiles} we see that
while at $t=1.5$ Gyr all bars had the classic shapes of $A_2$ profiles, increasing right from $R=0$, at $t=2$ Gyr the
$A_2$ for simulations S1 and S2 are below 0.1 up to $R=3$ kpc and only increase at larger $R$. The calculation of the
phase of the bar mode confirms that it is not constant with radius (as is characteristic for bars) any more and in this
period the bar is replaced by spiral arms. Soon after however, at around $t=2.2$ Gyr the bar starts to rebuild itself
in both simulations and is preserved until the end. The same happens in simulation S3 but slightly earlier,
around 1.85 Gyr from the start of the simulation.

\begin{figure}
\gridline{\fig{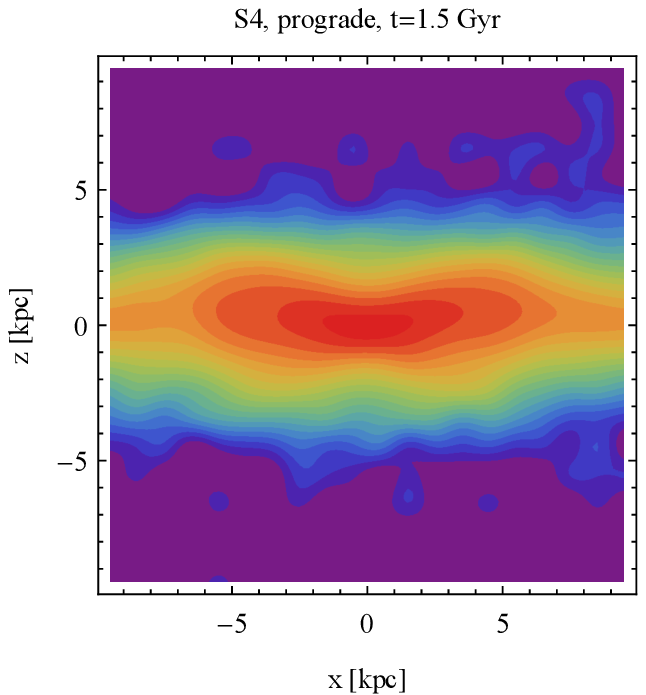}{0.4\textwidth}{}}
\vspace{-0.04\textwidth}
\gridline{\hspace{0.055\textwidth}
	  \fig{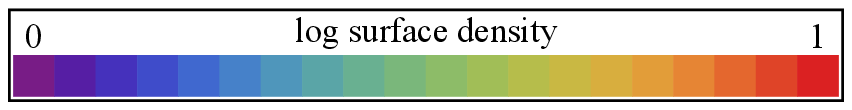}{0.3\textwidth}{}}
\vspace{-0.03\textwidth}
\caption{Example of the buckling instability occurring at $t=1.5$ Gyr from the start of the evolution in simulation S4
for the galaxy on prograde orbit. The image shows the surface density distribution of the stars in the edge-on view
(perpendicular to the bar). The density scale has been normalized to the maximum density of the stars in the center.}
\label{examplebuckling}
\end{figure}

\begin{figure}
\gridline{
	\fig{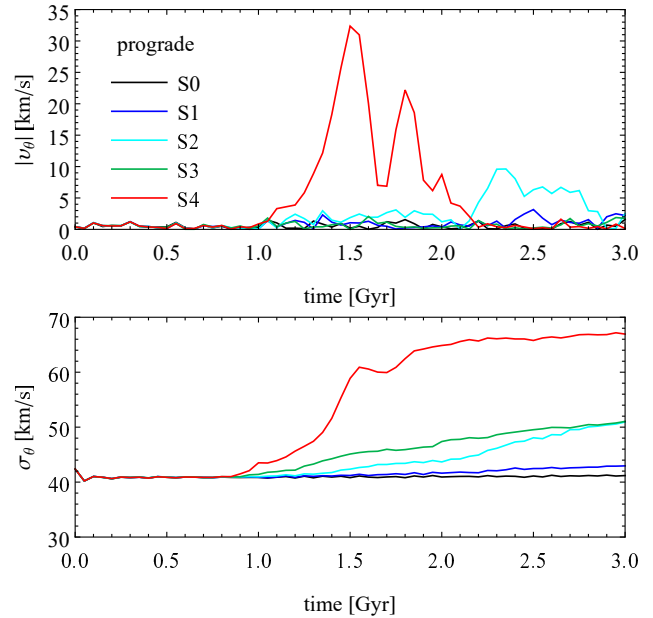}{0.46\textwidth}{}
	}
\vspace{-0.03\textwidth}
\caption{The evolution of different measures of buckling of the stellar component in time in simulations S0-S4
for galaxies on prograde orbits. Results for simulations
with different initial configurations are denoted with different colors, as indicated by the legend in the upper panel.
The upper panel shows the evolution the absolute value of the mean velocity along the spherical coordinate $\theta$,
$|v_\theta|$, and the lower one the corresponding velocity dispersion, $\sigma_\theta$. The measurements were performed
for stars within 7 kpc from the center of the galaxy.}
\label{buckling}
\end{figure}

In the case of S4, the bar starts to weaken as well, but even earlier, already around $t=1.5$ Gyr. In this case the
weakening is related to the buckling instability that is strongest around this time.
Figure~\ref{examplebuckling} shows the edge-on view of the stellar component in this simulation at this time.
Clearly, the bar is distorted out of the initial disk plane, as expected during strong buckling. We have verified this
by calculating kinematic buckling diagnostic in terms of the absolute value of the mean velocity along the spherical
coordinate $\theta$, $|v_{\theta}|$, of stars within a fixed radius of 7 kpc, where $\theta$ measures the angle with
respect to the disk plane. As illustrated in the upper panel of Figure~\ref{buckling}, the value of this
quantity starts to increase almost immediately after the bar formation, reaches a maximum at $t=1.5$ and decreases down
to zero again around $t=2.2$ Gyr, signifying the end of buckling, at least in the inner part of the galaxy. The bar
then rebuilds itself and grows in length (but not strength) until the end of the simulation at $t=3$ Gyr. As expected,
the buckling also results in the thickening of the bar (see the evolution of $c/a$ in the second panel of
Figure~\ref{shape}) and an increase of the velocity dispersion along $\theta$, $\sigma_{\theta}$ (see the
lower panel of Figure~\ref{buckling}), but no clear boxy/peanut shape forms.

We note that much weaker buckling can be seen also in simulation S2, but it starts around $t=2.1$ Gyr and therefore
cannot be responsible for the temporal dissolution of the bar around $t=2$ Gyr. In fact, for this galaxy the buckling
lasts until the end of the simulation, so the galaxy is still distorted a bit out of the disk plane at $t=3$ Gyr.
Interestingly in simulation S3 no buckling signal is seen in $|v_\theta|$, which remains at the zero level at
all times, but still $\sigma_\theta$ increases steadily with time, as does $c/a$. This means that some stellar orbits
depart from the disk plane but there is no consistent movement up or down that would lead to the uniform distortion of
the bar. The occurrence of buckling is known to reduce the bar strength, and it indeed does in the case of S4, as
proven by the drop of the mean $A_2$ value at $t=1.5$ Gyr in the lower left panel of Figure~\ref{shape}. In the case of
S3 the bar does not undergo buckling and is able to grow steadily unhindered. We believe this is the reason why the
bar in the S3 case ends up stronger and longer than in S4.

\begin{table}
\begin{center}
\caption{Properties of the bars after 3 Gyr of evolution for galaxies on prograde orbits}
\begin{tabular}{lccccc}
\hline
\hline
Simulation  & $A_{2,{\rm max}}$ & $a_{\rm b}$ & $\Omega_{\rm p}$ & $R_{\rm CR}$ &
$R_{\rm CR}/a_{\rm b}$\\
            &   &   [kpc]     & [km/s/kpc]       &  [kpc]    &    \\
\hline
\ \ \ \ S0 &   0.12     & 5.3    & 16.2   & 12.8  & 2.4 \\

\ \ \ \ S1 &   0.15     & 3.9    & 16.0   & 12.9  & 3.3 \\

\ \ \ \ S2 &   0.27     & 5.0    & 15.0   & 13.6  & 2.7 \\

\ \ \ \ S3 &   0.52     & 7.5    & 14.0   & 14.2  & 1.9 \\

\ \ \ \ S4 &   0.46     & 6.5    & 11.5   & 16.5  & 2.5 \\
\hline
\label{barproperties}
\end{tabular}
\end{center}
\end{table}

Table~\ref{barproperties} summarizes the properties of the bars formed in galaxies on prograde orbits after 3 Gyr of
evolution. For completeness and as a reference, in the first row we also list the same properties calculated
for the isolated case S0, although here the distortion is a very weak bar, just starting to form. Columns 2-6 of
the Table list the value of the first maximum of the bar mode profile $A_{2,{\rm max}}$ (see Figure~\ref{a2profiles}),
the length of the bar $a_{\rm b}$, the pattern speed $\Omega_{\rm p}$, the corotation radius $R_{\rm CR}$ and the ratio
$R_{\rm CR}/a_{\rm b}$. The pattern speeds were calculated from the angular variation of the major axis of the stellar
component in time and were found to be approximately constant in time. The corotation radii were computed by comparing
the circular frequency and the pattern speeds of the bars in the final simulation outputs. The length of the bar was
estimated as the cylindrical radius where the value of $A_2 (R)$ after reaching a maximum drops to $A_{2,{\rm max}}/2$
or, if this does not happen, a radius where a clear first minimum of the profile occurs and at the same time the phase
of the bar mode remains constant with radius. We note that the determination of the bar length is difficult in all the
interacting cases because of the presence of spiral arms and it is not exactly clear where the bar ends and the spiral
arms begin. In all cases the ratio $R_{\rm CR}/a_{\rm b}$ is of the order of 2 or higher indicating that our bars are
slow. We note that there is an average trend of stronger bars having lower pattern speed, a relation known to exist in
all types of bars. The similar trend in $R_{\rm CR}/a_{\rm b}$ is much less clear for our tidally induced
bars but the values of this parameter are very uncertain due to the difficulty in measuring the bar length.

In all our galaxies that evolved on prograde orbits additional morphological structures, besides the bar, are present.
These most often take the form of tidally induced spiral arms, visible in Figure~\ref{a2profiles} as high $A_2$ values
beyond radii 5-7 kpc. These grand-design, two-armed spirals originate from the stars stripped from the
galaxies at the time of the encounter and forming extended tidal tails immediately after. The arms later wind up and
weaken so that they are substantially diminished towards the end of the evolution. We note that these structures,
visible as inclined yellow stripes in the outer radii in Figure~\ref{a2modestime}, are more persistent and regular in
the case of the weakest interaction S1 in comparison to the stronger-interaction cases S2 and especially S3 and S4 where
they become more jagged. This is also to some extent visible in the upper panels of Figure~\ref{a2profiles}, where the
hierarchy of $A_2$ strength in the spiral arms is reversed with respect to the strength of the bars: the spirals are
strongest for S1 which produces the weakest bar. Interestingly, the tidally induced spiral arms also form in the
retrograde galaxy in the strongest interaction, S4, as shown by the red line that dominates at all times in the lower
panels of Figure~\ref{a2profiles}. The structure of these arms is however different, they are shorter and broader than
the ones induced in the prograde galaxies.

\section{Why are prograde and retrograde encounters different?}

In this section we explain why the outcomes of the interaction are different for galaxies on prograde and retrograde
orbits, in particular why tidally induced bars form only in the prograde cases. For this purpose we restrict our
considerations to the case of the weakest interaction in simulation S1 as an example, keeping in mind that stronger
interactions will obviously lead to the same results, only more pronounced, thus proving the trend in general.
In the following figures we will mark the results for the prograde galaxy in S1 in red and the retrograde one in green,
while the reference case of the same galaxy evolved in isolation (S0) will be shown in black.

\begin{figure}
\gridline{\fig{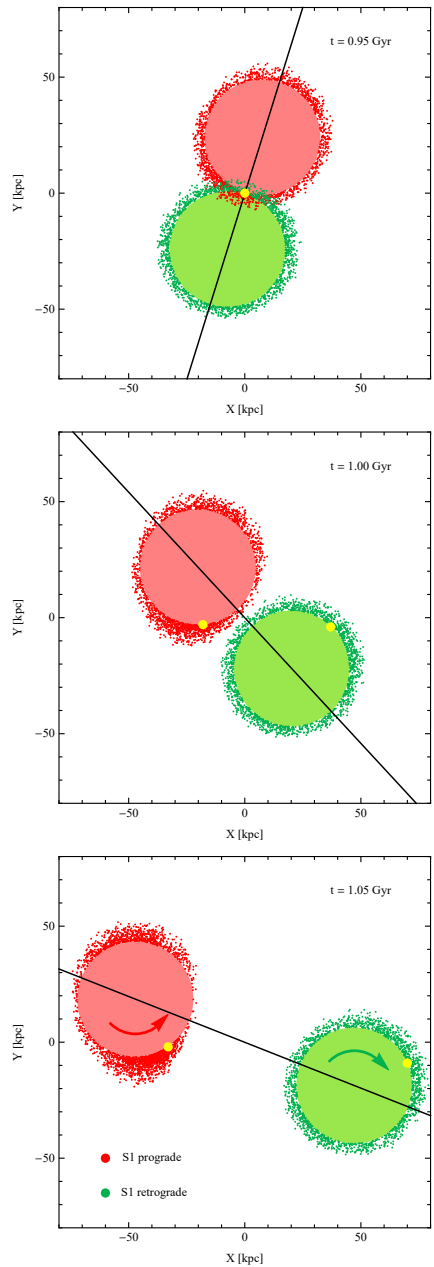}{0.31\textwidth}{}}
\vspace{-0.03\textwidth}
\caption{Face-on views of the interacting galaxy disks in simulation S1 close to the pericenter passage. The three
panels from top to bottom show the disks at times $t=0.95$, $t=1$ and $t=1.05$ Gyr from the start of the simulation.
The red disk is on the prograde orbit and the green disk on the retrograde one with the arrows in the lower panel
indicating the rotation direction of the disks. The inner parts of the galaxies ($R < 25$ kpc) were masked with disks
of uniform colors, while the dots show stars at $R > 25$ kpc. The straight black lines join the centers of the galaxies
thus indicating the direction of the tidal force. Yellow circles show the motion of an example star belonging to each
of the galaxies assuming it moves on an undisturbed circular orbit with velocity $\sim 200$ km s$^{-1}$.}
\label{overview}
\end{figure}

It is instructive to first look directly at the simulation snapshots near the pericenter passage. Three stages of the
encounter showing the stellar components of the interacting galaxies are displayed in Figure~\ref{overview}. At
$t=0.95$ Gyr (upper panel) the galaxies are seen just before the closest approach, the two following panels show the
situation 0.05 and 0.1 Gyr later, just after the pericenter. The solid black lines join the centers of the galaxies and
show the direction of the tidal force. The essence of the difference between the effect of the interaction on the
prograde versus the retrograde galaxy can be grasped by looking at the motion of an example yellow stars belonging to
each of the galaxies. We chose the stars to lie in the same location in the upper panel, just on the line connecting
the centers of the galaxies, at $R=25$ kpc from each of them. Assuming that the stars move on unperturbed circular
orbits with a velocity $\sim 200$ km s$^{-1}$, they move along the direction of rotation in each disk, as shown in the
next panels. Since the star in the prograde disk moves in the same direction as the perturber, in the next two stages
it remains close to the direction of the tidal force (black line). On the other hand, the star in the retrograde disk
moves away from this direction.

Thus the tidal force acts on the stars in the prograde galaxy for a longer time
than on the corresponding stars of the retrograde disk. To put it more precisely, tidal forces operate on the same
subsample of stars in the prograde disk for a longer time, while in the retrograde one different stars along the
circumference of the galaxy are affected in sequence. The result is the formation of elongated, narrower tidal
structures in the prograde galaxy while in the retrograde one only weak ring-like and concentric distortions are
present. The difference in the distribution of stars in the two galaxies is already visible right after the
interaction, in the lowest panel of Figure~\ref{overview}, and becomes more pronounced in time as the narrow tidal
streams become longer and evolve into spiral arms. These stronger perturbations propagate towards the center giving
rise to the formation of the bar.

\begin{figure}
\gridline{\fig{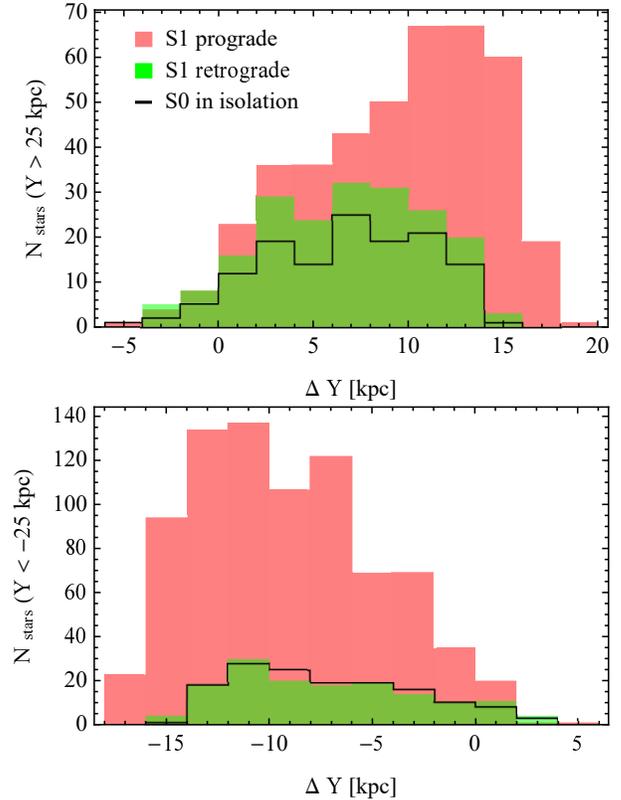}{0.44\textwidth}{}}
\vspace{-0.03\textwidth}
\caption{Histograms showing numbers of stars experiencing different displacements along the $Y$ axis, the direction of
the tidal force at the pericenter. The positions of the stars were measured along the $Y$ axis of the simulation box but
with respect to each galaxy center. The upper histogram is for stars having $Y > 25$ kpc (upper parts of the disks) at
$t=1.05$ Gyr that changed their position by $\Delta Y$ during the last 0.1 Gyr. The lower histogram is for stars in
the lower parts of the disks, with $Y < -25$ kpc. Red histograms refer to stars of the prograde galaxy, green ones to
stars of the retrograde galaxy and the black line ones to the galaxy evolving in isolation.}
\label{histograms}
\end{figure}

To illustrate the difference in the perturbations occurring in the prograde and retrograde galaxy in a more
quantitative way, in Figure~\ref{histograms} we show histograms of the number of stars experiencing a given shift in
their positions along the $Y$ axis of the simulation box, which is also the direction of the tidal force at the
pericenter. The positions of the stars were measured along the $Y$ axis, but with respect to the center of a given
galaxy. As in the previous figure, the red color codes the results for the prograde galaxy and the
green one for the retrograde disk. For comparison we also show in black the measurements for the galaxy evolved in
isolation. In the upper histogram we show the distribution of stars that at $t=1.05$ Gyr (the configuration shown in
the lowest panel of Figure~\ref{overview}) found themselves in the upper parts of the disks, $Y > 25$ kpc, and changed
their positions by $\Delta Y$ during the last 0.1 Gyr, i.e. the whole time period illustrated in Figure~\ref{overview}.
The lower histogram shows the same for stars in the lower parts of the disks, with $Y < -25$ kpc.

Let us first look at the results for the galaxy in isolation, shown by the black line histograms. Note that the two black
histograms in the upper and lower panel are in fact very similar and only differ in appearance because of the different
scale of the number of stars in each panel (vertical axis). This is expected since for an unperturbed disk, the stars
that find themselves now at their extreme distance from the galaxy center along the $Y$ axis should have travelled the
same distance on average, no matter on which (upper or lower) side of the galaxy they are. For a star moving with the
velocity of $\sim 200$ km s$^{-1}$, as is typical for our model, on a circular orbit of radius $R=25$ kpc this distance
should be around $\Delta Y = \pm9$ kpc, which is indeed the value where the maxima of the distributions occur.
Obviously, not all stars are at their maximum $Y$ and not all follow exactly circular orbits (because of the velocity
dispersion inherent in the models) so the values of $\Delta Y$ show considerable scatter.

\begin{figure}
\gridline{\fig{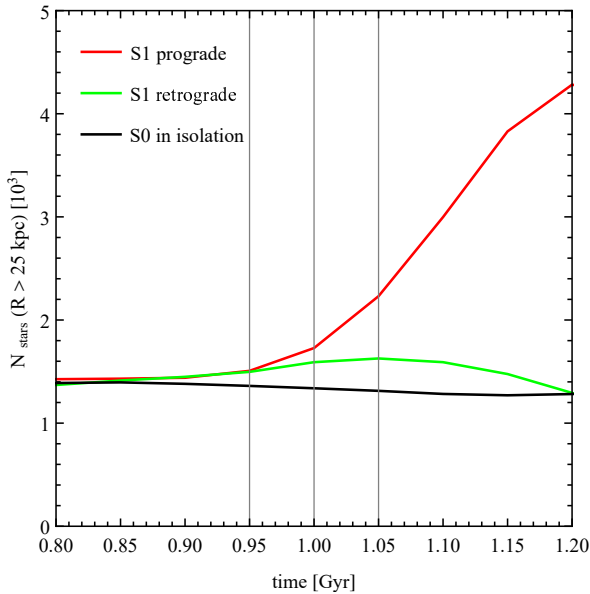}{0.43\textwidth}{}}
\vspace{-0.03\textwidth}
\caption{The number of stars outside the radius of $R = 25$ kpc as a function of time for the prograde (red) and
retrograde (green) galaxy in simulation S1 and the isolated galaxy (black). Thin vertical gray lines indicate the
times of the snapshots shown in Figure~\ref{overview}. The pericenter of interaction S1 occurs at $t=0.97$ Gyr.}
\label{outerstars}
\end{figure}

For the galaxy on the retrograde orbit (green histograms in Figure ~\ref{histograms}) the numbers of stars
with given shifts $\Delta Y$ turn out to be the same as for the isolated galaxy on the side more distant from the
perturber (negative $Y$, lower histogram) or slightly higher on the side closer to the perturber (positive $Y$, higher
histogram). Such asymmetry of the tidal effects is expected from theoretical considerations \citep[e.g.][]{DOnghia2010}.
For the prograde galaxy (red histograms) the effect of tidal forces is much stronger: there are many more stars shifted
to the outskirts of the galaxy and their typical shifts are larger, with the maxima of the distributions falling around
$\Delta Y = \pm12$ kpc. Here again strong asymmetry is present with about twice more stars stripped on the side closer
to the perturber (now negative $Y$, lower histogram) than the more distant one (now positive $Y$, higher histogram).

Since the stripping does not occur solely along the $Y$ axis, in order to grasp the outflow of stars in all directions,
in Figure~\ref{outerstars} we plot the number of stars outside the radius of $R = 25$ kpc as a function of time near
the pericenter passage. As before, the red and green lines show measurements for the prograde and retrograde galaxy
respectively in simulation S1 and the black one refers to the galaxy in isolation. The three gray vertical lines
indicate the times of snapshots shown in Figure~\ref{overview}. Clearly, the results confirm what was shown in
Figure~\ref{histograms} for the shorter time period of 0.1 Gyr: the outflow of stars in much stronger for the prograde
galaxy than for the retrograde one. Although the retrograde galaxy is certainly more affected than the isolated one,
the stripped stars are accreted back shortly and later on oscillate around the level characteristic of the galaxy in
isolation. While some fluctuations are also present for the prograde galaxy at later times, there is always a
significant number of stars in the outer parts that form the extended tails and then evolve into spiral arms.

\section{Conclusions}

We studied the dynamical and morphological evolution of Milky Way-like galaxies in flybys where one galaxy was on
exactly prograde and the other one on exactly retrograde orbit. The initial configurations differed by the impact
parameter and the relative velocity of the galaxies. We found that significantly strong tidally induced bars form only
in galaxies on prograde orbits, while those on retrograde ones are affected very little, with distortions similar to
those developed by the same galaxy evolved in isolation.

We therefore confirm the earlier findings of \citet{Lokas2015} and \citet{Lokas2016} where the formation of tidally
induced bars was investigated in the context of galaxies orbiting a bigger host, i.e. dwarfs around the Milky Way or
normal-size galaxies in clusters. Both these studies found that bars formed only in galaxies on prograde orbits. Our
results agree with those of \citet{Lang2014} who studied encounters of galaxies of similar mass and found that tidal
bars can only form in prograde orientations as we do here.

Our conclusions are however different from those of \citet{Martinez2017} who found no significant differences between
bars formed on prograde and retrograde orbits. The discrepancy is probably due to the use of the impulse
approximation by \citet{Martinez2017}. The impulse approximation applies only if the
encounter time is much shorter than the internal crossing time of a galaxy \citep{Binney2008}. While it can be
applied to modify the stellar velocities instantly, the impulse approximation in the form used by
\citet{Martinez2017} does not distinguish between the prograde and retrograde encounters.

\citet{Martinez2017} claim to have tested the robustness of the impulse approximation by a simulation
where the perturber is modelled by a set of particles and the encounter is prograde. However, they did so
for 1) a galaxy that forms a strong bar in isolation and the interaction, whether modelled by the impulse approximation
or a real simulation, has very little effect on the properties of the bar; 2) a galaxy that does not form a bar at all,
neither in isolation, nor in interactions. Thus these experiments cannot be considered as a valid proof of the
applicability of the approximation in this context.

As discussed by
\citet{DOnghia2010}, the impulse approximation can be extended to account for the difference between prograde and
retrograde encounters by including higher order terms in the derivatives of the potential involved in the calculation
of velocity increments. Only these higher order terms include the dependence of the result on the relative orientation
between the position vector of the star in the galaxy and the direction to the perturber as well as their dependence on
time.

In their analytical calculations
\citet{DOnghia2010} assume that the stars do not depart from their original trajectories during the encounter and all
move on exactly circular orbits. As we have demonstrated in the previous section, neither of these conditions is
fulfilled in realistic $N$-body simulations so we cannot expect quantitative agreement between the
predictions of such formulae and $N$-body results. The analytical results allow however to grasp the difference between
prograde and retrograde encounters and predict a quasi-resonant response for stars with circular frequency similar to
the orbital frequency of the perturber on prograde orbits. The resonance is broad in a sense that the frequencies do
not have to match exactly and can differ by a factor of a few, as is the case in the simulations presented here.

\section*{Acknowledgments}

This work was supported in part by the Polish National Science Centre under grant 2013/10/A/ST9/00023.
Helpful comments from an anonymous referee are kindly appreciated.
\software{GADGET-2 \citep{Springel2001, Springel2005}}.

\end{document}